# Nonlocal communication with photoinduced structures at the surface of a polymer film

**Régis Barille, Sylvie Dabos-Seignon and Jean-Michel Nunzi**

*Laboratoire POMA UMR 6136 - Université d'Angers, 2 Boulevard Lavoisier, 49045 Angers, France*
*regis.barille@univ-angers.fr, jean-michel.nunzi@univ-angers.fr*

**Sohrab Ahmadi-Kandjani**

*Laboratoire POMA UMR 6136 - Université d'Angers, 2 Boulevard Lavoisier, 49045 Angers, France*
*on leave from RIAPA, University of Tabriz, 51664 Tabriz, Iran*

**Ewelina Ortyl and Stanislaw Kucharski**

*Institute of organic and polymer technology, Wroclaw Technical University, 50-370, Wroclaw, Poland*
*Stanislaw.Kucharski@pwr.wroc.pl*

**Abstract:** Nonlocal communication between two laser light beams is experimented in a photochromic polymer thin films. Information exchange between the beams is mediated by the self-induction of a surface relief pattern. The exchanged information is related to the pitch and orientation of the grating. Both are determined by the incident beam. The process can be applied to experiment on a new kind of logic gates.

## 1. Introduction

There is still an open question concerning the possibility of using simple optics in replacement of the next generations of computers or at least to perform purely optical logic operations and calculations. We show in this work that it is possible to implement an optical network and transfer information from one point to another point without overlapping of the light beams. Logical operations between the beams can also be performed. The optical-network model of computation consists in many elementary processing nodes (neurons), communicating each other with interconnection weights [1,2]. Associated with each node is an activation level which is transformed into a weighted sum of the activity levels of all the other nodes [3]. The node in our experiment is a surface induced relief grating (SRG) whose input is both the pitch and the orientation coded when the network is operated in the static mode. The grating which results from a photodynamic polymer transport process is produced in photoactive polymer thin films with one single laser beam and shows regularly spaced surface relief structures [4–6]. The grating spacing and orientation can be controlled by the incidence angle and polarization of the incident laser beam, respectively [7].

All the networking nodes can be activated simultaneously or sequentially by a laser beam with sufficient exposure time per node. The neuron output is then the acquisition of a polarization state and grating pitch, and subsequent light diffraction. The network in our system is a set of multiple SRG. We implement and test such system using the nonlocal self organization aspect of SRG under incoherent light power. Indeed, we reported previously in reference [8] that when we send a low intensity coherent seed beam simultaneously with a large intensity incoherent pump beam on the photopolymer surface, a SRG is induced and propagates all over the incoherent beam area, outside of the coherent beam: it bears all the features of the SRG induced by a large intensity coherent beam. We believe that our simple experimental optical processor could be useful to implement and test new ideas relevant to neuronal photonics.

## 2. Materials and methods

Samples are polymer films made from a highly photoactive azobenzene derivative containing heterocyclic sulfonamide moieties: 3-[{4-[(E)-(4-{[(2,6-dimethylpyrimidin-4-yl) amino] sulfonyl}phenyl) diazenyl]phenyl}-(methyl)amino]propyl 2-methylacrylate [9]. Thin films on glass substrates were prepared by spin-coating of the polymer from THF solutions with a concentration of 50 mg/ml. Thickness measured with a Dektak-6M Stylus Profiler was around 1μm. The $\lambda$ = 476.5 nm laser line of a continuous argon ion laser is used to excite the azo-polymer close to its 438 nm-absorption maximum. Absorbance at the working wavelength is 1.6. Polarization direction of the laser beam is varied using a half-wave plate. Our experimental set-up is sketched in Fig. 1. Laser power is 500 mW. The beam width is increased to 5.8 mm by a two lenses afocal system. The beam is then divided in the two equal arms of a Mach-Zehnder. In the first arm, the beam is reduced to a width $w_0$ = 750 μm by another afocal and is divided into two equal power spots. Power of each beam is attenuated down to 0.07 mW. These two narrow beams are the signal beams. The second one is the pump beam. The pump beam is made spatially incoherent by focusing the laser beam through a diffuser. It can also be made temporally incoherent with the signal beams using an additional delay line with retardation path larger than the 5.7 cm-coherence length of our laser. An additional hot-air gun is used as a turbulent layer generator and is inserted along the beam path in order to insure total incoherence and depolarization. The diffuser and the turbulence generator introduce a random phase which varies much faster than the time constant for grating formation in the sample. The signal and pump beams are spatially overlapped at the exit of the Mach-Zehnder, with the thinner signal beams fixed in the pump beam central region. They illuminate the polymer sample. Pump beam power is 87 mW.

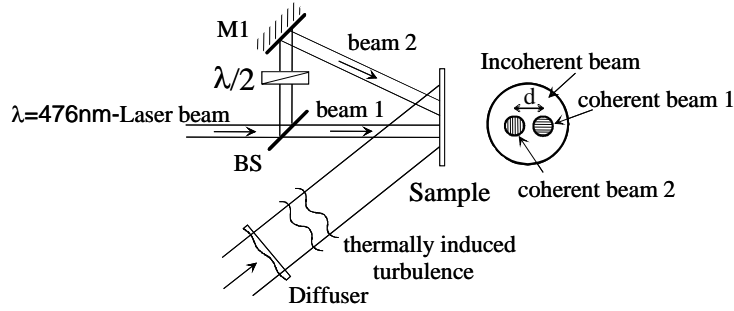

Fig. 1. Experimental set-up. Beam 1 is at normal incidence on the sample. Beam 2 is at a 32°-incidence angle. With respect to the incidence of beam 2, the polarization is TM for beam 1 and approximately TE for beam 2. BS: beam splitter; M1: mirror; λ/2: half wave plate.

## 3. Experiments and discussion

We first checked the material response for the coherent signal beams alone, without incoherent pump beam. In this test, the signal beams are not attenuated, their power is 0.6 mW. The two beams are sent on the sample (Fig. 1) and the SRG is recorded after one hour illumination. The distance between the edges of the two $w_0 = 750$ μm spots is fixed to $d = 920$ μm. One of the beams (beam 1) has 32°-incidence angle and the other (beam 2) is normal to the sample surface. Both the height and pitch of the grating were retrieved after illumination with a contact-mode AFM. For beam 1, the grating pitch is $\Lambda_1 = 880$ nm and for beam 2, $\Lambda_2 = 590$ nm, both with a grating amplitude of 150 nm ± 20 nm. These pitches are in agreement with the values calculated by first order diffraction at an angle θ. In both cases the pitch is uniform in one direction and the two dimensional Fourier transforms in Fig. 2(a) show only one single spatial frequency component $\nu_x$ or $\nu_y$. This shows that the two beams do not interact.

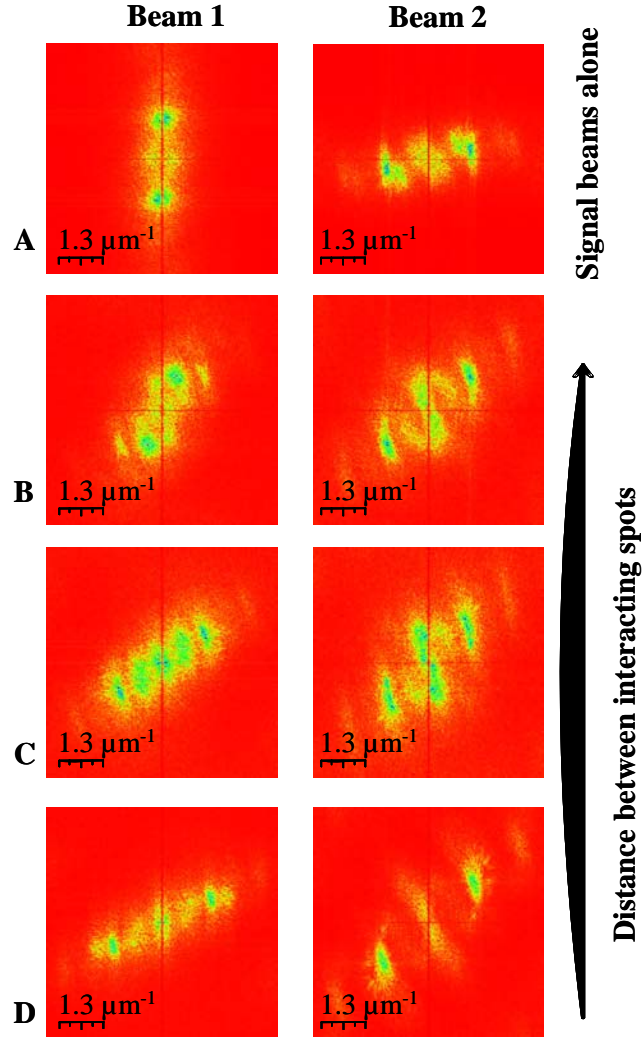

Fig. 2. Two dimensional Fourier transform of the two SRG spots (A) under the two signal beams without the incoherent pump. The same with the incoherent pump overlapping with the signal beams separated by a distance $d = 920$ μm (B), $d = 650$ μm (C) and $d = 450$ μm (D). The SRGs under beams 1 and 2 have a pitch $\Lambda_1 = 880$ nm and $\Lambda_2 = 590$ nm, respectively. The Fourier transform reflects the far field diffraction pattern of the SRG.

In a second step the 87 mW incoherent pump beam overlaps with the two 0.07 mW coherent signal beams. We check SRG construction for three different horizontal separations between the edges of the spots: 920, 650, and 450 μm, corresponding to a ratio $w_0/d$ of 1.6, 1.15, and 0.8 respectively. Figure 2 shows the two dimensional Fourier transforms of the SRG below beams 1 and 2. When the two spots get closer, we see that the spatial frequency spectrum is enlarged. For the SRG below beam 1, the rotation of its frequency spectrum corresponds to the mutual interaction with beam 2. We see the same rotation in a reverse direction for beam 2, the two grating orientations becoming almost the same. When the two spots become closer (Fig. 2(d)), the two spatial frequency spectra become very similar. This shows that the two beams interact, the information on polarization is exchanged first at large distance and the information about the pitch is mixed, with appearance of new frequencies at shorter distance. Surface profile and power spectral densities (PSD) of these images are represented in Fig. 3 for two cross-sections in the vertical (blue lines) and horizontal (red lines) directions and for ratios $w_0/d$ of 1.7 and 0.8. For each beam, the spatial frequency corresponding to the original pitch is conserved, with additional frequencies appearing when the separation distance is reduced. For the spot under beam 1 the horizontal frequency $v_x = 1/\Lambda_2$ appears in the power spectrum for the shortest separation. In the same way, for the spot under beam 2 the vertical frequency $v_y = 1/\Lambda_1$ appears in the power spectrum for the shortest separation. The numerical criterion for appearance of a new spatial frequency is chosen arbitrarily in our experiment as being a peak emerging above 50% in the normalized PSD spectrum. That is the way the mutual influence between the 2 beams takes place.

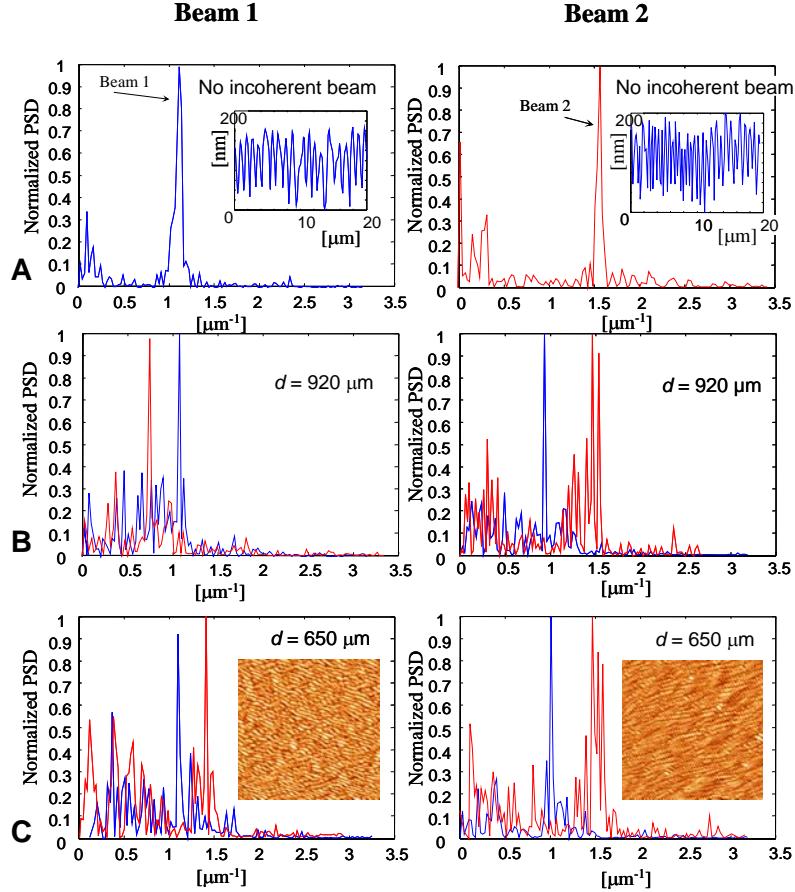

Fig. 3. Power spectral density for different distances between the spots. PSD is measured for the vertical (blue lines) and horizontal (red lines) cross-sections of the SRG patterns under beam 1 (left) and 2 (right). Without coupling by the incoherent beam (A), coupled with 920 μm distance (B) and coupled with 650 μm distance (C). Inset in (A) shows the surface profile. Inset in (C) shows the AFM scan of a 20×20 μm² region below the signal beams.

A last experiment was performed with the half-wave plate placed on the path of beam 1 and its polarization fixed at 45°. Beam 2 polarization is vertical as given by the laser. The distance between the spots is reduced to $d = 180$ μm. A photodiode is set to measure the first order which would be diffracted from the self-induced SGR by a beam 2 with horizontal polarization. We must recall that the direction of diffraction follows the polarization direction [7], as shown in Fig. 2(a). This signal, shown in Fig. 4, is the transfer of polarization from beam 1 to beam 2, associated with the new horizontal spatial frequency created beam 2 in Fig. 3(c). This new spatial frequency induces the horizontal diffraction which is detected by the photodiode. The signal in Fig. 4 saturates after 100 minutes of illumination, when the information transfer is complete.

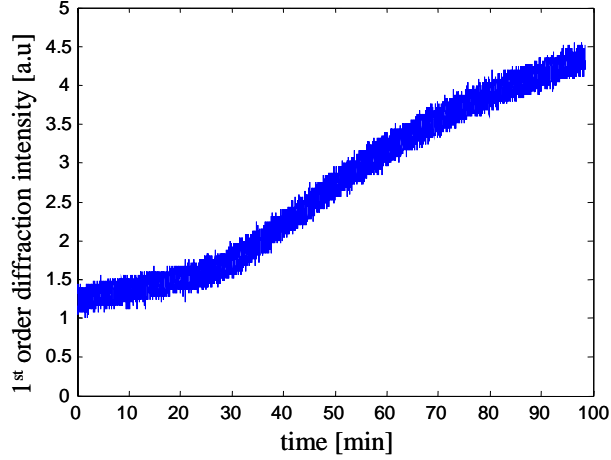

Fig. 4. First order diffraction intensity as a function of time for the coherent beam 2. Beam 1 polarization is fixed at 45° and beam 2 polarization is vertical. The photodiode is set to measure the first order of diffraction from beam 2 in the horizontal direction. The signal detected is a signature of the polarization transfer from beam 1 to beam 2.

Different nodes can be lighted up on the surface below the incoherent beam,. Associated with each SRG node $i$ is its spatial Fourier spectrum described by a function $y_i^{(n)} = f(x_i^{(n)})$, where $x_i$ is the activation level created by the weighted sum of the activity parameters of the other nodes: $x_i^{(n)} = \sum_i \omega_{ij} y_j^{(n-1)}$. As shown in Fig. 3, the weights $\omega_{ij}$ are function of the distance $w_0/d$ between the nodes. The number of iterations $n$ is 1 in our experiment. We implemented two basic logic operation in our simple network using this neuronal principle. The results of the two Boolean operations OR and AND are given in tables 1 and 2, respectively. In table 1, we consider the pitch of the SRG under beam 1 as test parameter $y_1$. We assign the value 1 to the pitch of the SRG under the beam 1 which arrives alone at normal incidence. Any other SRG pitch will be assigned the value 0. The pitch $y_i^{(0)}$ of the SRG is addressed by the beam incidence. The results in table 1 show that we implement an OR gate in this way.

Table 1. Implementation of an OR logic gate with pitch $\Lambda_i$ as test parameter.

| $y_1^{(0)} \equiv \Lambda_1$ | $y_2^{(0)} \equiv \Lambda_2$ | $y_1^{(1)}$ |
|---|---|---|
| 1 | 1 | 1 |
| 0 | 0 | 0 |
| 1 | 0 | 1 |
| 0 | 1 | 1 |

In table 2, we consider the orientation of the SRG under beam 1 as test parameter $y_1$. We assign the value 1 to a vertical orientation of the SRG under the beam 1 which arrives alone with vertical polarization. Any other orientation will be assigned the value 0. The orientation $y_i^{(0)}$ of the SRG is addressed by the beam polarization. The results in table 2 show that we implement an AND gate in this way. We see that our simple network can perform simple logic operations. More complicated operations can be achieved if more signal beams interact within the incoherent beam at the polymer surface.

Table 2. implementation of an AND logic gate with polarization $P_i$ as test parameter.

| $y_1^{(0)} \equiv P_1$ | $y_2^{(0)} \equiv P_2$ | $y_1^{(1)}$ |
|---|---|---|
| 1 | 1 | 1 |
| 0 | 0 | 0 |
| 1 | 0 | 0 |
| 0 | 1 | 0 |

## 4. Conclusion

We have implemented a very simple neurocomputer where the individual neuron is the self-induced SRG below a coherent signal beam. In this experiment, the network is configured in the static mode and each neuron is configured by the photoinduced polymer transport process. The neurons interact through the self-induced SRG which dissipate the strong incoherent pump beam. The learning process of each neuron consists in the acquisition of the pitch and polarization parameters. The result of the activation is processed from the weighted sum of the activity levels of the individual neurons. This basic implementation can be extended to a more complex network structure. For instance, a large number of nodes can be implemented in a single polymer film under incoherent illumination. The neurons can also interact between several cascaded planes as each individual neuron bears an underlying SRG which diffracts efficiently in privileged directions [7]. The process is partly reversible: work is now in progress to exploit the dynamic reconfiguration of the network (more than $n = 1$ iteration) which is permitted as long as saturated growth of the SRG is not achieved [7]. Work is also in progress to improve the time response of the system using faster responsive polymers. We see that our simple experiment on optical processing is relevant to the implementation and evaluation of new concepts related to neuronal photonics.